\documentclass[authoryear,preprint,review,11pt, 5p,times,twocolumn]{elsarticle}

\usepackage{graphicx}
\usepackage{subfig}
\usepackage{amsmath}
\usepackage{amsfonts}
\usepackage{amssymb}
\usepackage{url}

\journal{Social Networks}

\begin{document}
		\begin{frontmatter}
		\title{ Weighted reciprocity in human communication networks\tnoteref{a}}		
		\tnotetext[a]{Research was sponsored by the Army Research Laboratory and was accomplished in part under Cooperative Agreement Number W911NF-09-2-0053,   by the Defense Threat Reduction Agency (DTRA) grant HDTRA 1-09-1-0039 (Anthony Strathman and Zolt\'{a}n Toroczkai), and by the National Science Foundation (NSF) grant BCS-0826958. The views and conclusions contained in this document are those of the authors and should not be interpreted as representing the official policies, either expressed or implied, of the Army Research Laboratory or the U.S. Government. The U.S. Government is authorized to reproduce and distribute reprints for Government purposes notwithstanding any copyright notation hereon.  Special acknowledgments go to Albert L\'{a}szl\'{o} Barab\'{a}si for providing the source data.} 
				
		\author[ad1,ad4]{Cheng Wang}
		\author[ad2,ad4]{Anthony Strathman}
		\author[ad1,ad4]{Omar Lizardo\corref{corr1}}
		\author[ad1,ad4]{David Hachen}
		\author[ad2,ad3,ad4]{Zolt\'{a}n Toroczkai}
		\author[ad3,ad4]{Nitesh V. Chawla}
		\address[ad1]{Department of Sociology, University of Notre Dame, 810 Flanner Hall, Notre Dame, IN, 46556}
		\address[ad2]{Department of Physics, University of Notre Dame, 225 Nieuwland Science Halll, Notre Dame, IN, 46556}
		\address[ad3]{Department of Computer Science and Engineering, College of Engineering, University of Notre Dame, 384 Fitzpatrick Hall, Notre Dame, IN, 46556}
		\address[ad4]{Interdisciplinary Center for Network Science and Applications (ICeNSA), 384E Nieuwland Science Hall, Notre Dame, IN 46556}		
		\cortext[corr1]{Corresponding author at \texttt{olizardo@nd.edu}.}
	
		\begin{abstract}
			In this paper we define a metric for reciprocity---the degree of balance in a social relationship---appropriate for weighted social networks in order to investigate the distribution of this dyadic feature in a large-scale system built from trace-logs of over a billion cell-phone communication events across millions of actors.  We find that dyadic relations in this network are characterized by much larger degrees of imbalance than we would expect if persons kept only those relationships that exhibited close to full reciprocity.  We point to two structural features of human communication behavior and relationship formation---the division of contacts into strong and weak ties and the tendency to form relationships with similar others---that either help or hinder the ability of persons to obtain communicative balance in their relationships.  We examine the extent to which deviations from reciprocity in the observed network are partially traceable to these characteristics.
		\end{abstract}
\end{frontmatter}

\section{Introduction}
Reciprocity has been recognized to be one of the most important properties of the connections linking entities in networked systems \citep{garlaschelli_loffredo04, boccaletti_etal06, wasserman_faust94, zamora-lopez_etal08, skvoretz_agneessens07}.  The study of dyadic reciprocity began in the sociometric and social network analysis tradition as a way to characterize the relative behavioral or cognitive ``balance'' in social relationships \citep{hallinan78, hallinan_hutchins80, hammer85, davis63, mandel00, krackhardt87, newcomb68}.  These studies defined reciprocity in a very simple---but fundamentally limited---way.  A dyad was reciprocal if both partners nominated one another as friends, or---in the tradition of ``balance theory'' \citep{heider58, newcomb61, newcomb79, davis79, doreian02}---if it was found that the relationship had the same valence (positive or negative) for both participants.  Dyads were viewed as non-reciprocal either when one partner reported considering the other one a friend or a close associate and the other did not, or if one partner displayed positive sentiments towards a partner who felt negatively towards him or her. The fundamental hypothesis of balance concerned a dynamic prediction: over time ties that were imbalanced were expected either to become balanced or to dissolve \citep{hallinan78, newcomb61, newcomb79, doreian02}.

This definition of ``reciprocity'' fit very well with the representation of social networks in early graph theory as consisting of binary (1,0) edges connecting two nodes \citep{wasserman_faust94}.  Analysts can then establish the level of reciprocity in the network via the so-called ``dyadic census.''  This presupposes a binary adjacency matrix A, where $a_{ij}=1$ if actor $i$ chooses actor $j$ as a neighbor and $a_{ij}=0$  otherwise.  Three types of dyads can then be defined: asymmetrical---sometimes also referred to as ``non-reciprocal'' ($a_{ij}=1$ and  $a_{ji}=0$ or   $a_{ij}=0$ and  $a_{ji}=1$), symmetrical ($a_{ij}=a_{ji}=1$) and null ($a_{ij}=a_{ji}=0$), otherwise known as the UMAN classification \citep{carley_krackhardt96}.  The phenomenon of dyadic reciprocity at the level of the whole network has been studied by comparing the relative distribution of asymmetric and mutual dyads in a graph \citep{mandel00, garlaschelli_loffredo04}.  We can consider levels of non-reciprocity to be high if the proportion of asymmetric dyads is larger than would be obtained by chance in a graph with similar topological properties (for instance a graph with the same number of nodes and edges).  This traditional definition of reciprocity has been extended and developed for the analysis of reciprocity in complex systems (social, technological, biological, etc.) organized as networks \citep{garlaschelli_loffredo04}.

The binary classification of dyads into three types misses one of the most important features of a dyadic relationship: the relative \emph{frequency} of contact between the two partners \citep{hammer85, eagle_etal08, eagle_etal09}.   This is a dimension of dyadic relationships that has always been considered crucial in previous treatments of the dynamics and static correlates of dyadic ties \citep{hammer85, marsden_campbell84, feld81, peay80}, but which has not been treated in depth in the existing literature, mainly due to lack of reliable behavioral data on repeated social interactions among humans in natural environments \citep{eagle_etal08}. Furthermore, it should also be clear that our intuitive notions of what reciprocity is requires information about the relative ``balance'' not of static mutual nominations or sentiments, but of repeated behavioral interactions, exchanges or flows in a dyad \citep{wellman_wortley90, borgatti05, borgatti_etal09}. This means that a more empirically accurate definition of reciprocity can only be obtained in the context of a \emph{weighted graph} \citep{barrat_etal04, yook_etal01, kossinets_watts06}---also referred to as ``valued graphs'' \citep{peay80, freeman_etal91, yang01}. In this representation, instead of a tie being thought of as simply being present or absent, the adjacency matrix is now defined by weights ($a_{ij}=w_{ij}$) which indicate the relative flow strength of the arc (e.g., the count of the number of interactions initiated by $i$ and directed towards $j$).

\begin{figure*}[t!]
	\centering
	\subfloat[\scriptsize Reciprocity as a result of outdegree matching]{
	\includegraphics[width=3.25in]{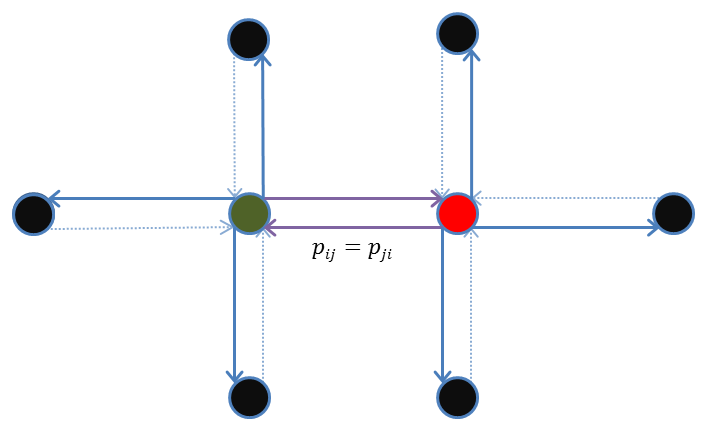}	
	\label{fig:config1}
		}
	\subfloat[\scriptsize Non reciprocity as a result of outdegree-mismatch.]{
	\includegraphics[width=2.25in]{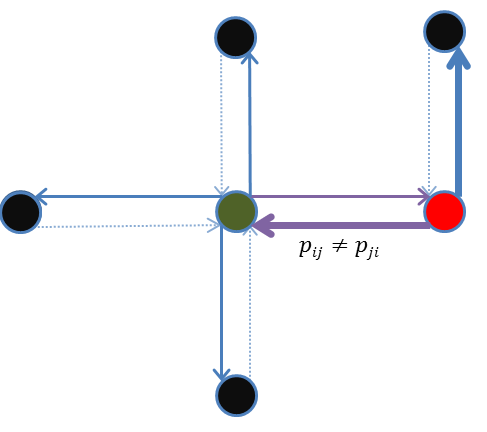}	
	\label{fig:config2} 
		}  \\
	\subfloat[\scriptsize Non-reciprocity as a result of directed weight-mismatch.]{
	\includegraphics[width=3.25in]{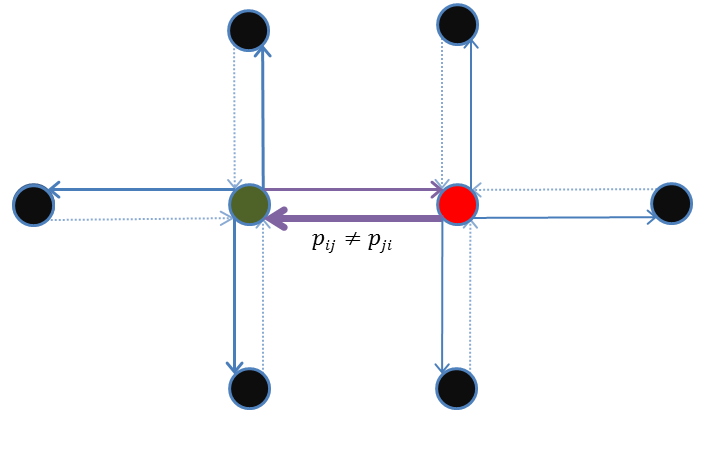}	
	\label{fig:config3}
		}
	\caption{Idealized local-structural scenarios producing different levels of dyadic reciprocity.}
	\label{fig:net-configs}
\end{figure*}

In this context, what have traditionally been considered ``non-reciprocal'' dyads---e.g., one partner in the dyads nominates the other but not vice-versa---\citep{carley_krackhardt96} can be better thought of as the one-way, that is, completely unbalanced limit of the interactions between the agents forming the dyad.  This is the case in which one partner in the relationship initiates all contact attempts and receives no reciprocation from the other member of the dyad.  Intuitively it is doubtful whether we can call this a relationship in the first place.   In the very same way, what have been traditionally conceived of as ``reciprocal'' (e.g. ``mutual'' or ``symmetric'') dyads can exhibit high levels of communicative imbalance with most of the interaction being one-way.  Consider for instance a dyadic relationship in which one partner is five times more likely to direct a communication towards the other person than the reverse.  It is not very intuitive to call this dyad ``reciprocal,'' but that is precisely what measurements methods that discard the information encoded in the weights do.

In sum, advancing research on dyadic reciprocity requires that we define dyadic reciprocity for weighted graphs.  In this paper we do that (section~\ref{sec:weighted}), while offering an empirical account of the distribution and correlates of dyadic reciprocity in a weighted social network built from trace logs of cell phone communications (section~\ref{sec:empirical}).  Finally, we compare (section~\ref{subsec:results}) the observed distribution to that obtained when we vary key dimensions of the networks that we argue are important topological and structural drivers of reciprocity.   We close by outlining the implications of our argument and results (section~\ref{sec:discussion}).

\section{Weighted reciprocity}
\label{sec:weighted}

\subsection{Weighted reciprocity metric}
We seek to define a measure of dyadic reciprocity that captures the degree of communicative imbalance a two-way relationship between two actors.   Consistent with the notion of reciprocity as balance, this measure should have the following properties: first, it should be at a minimum when the weight of the directed arc going from vertex $i$ to vertex $j$ approaches the weight of the directed arc going from vertex $j$ to vertex $i$.  Second, it should increase monotonically with the weight difference between the two directed arcs.  Third, it should normalize the weight difference to adjust for the fact that some persons are simply more communicative than others (they contact all of their partners more or less frequently). Finally, the measure should be the same irrespective of directionality ($R_{ij}=R_{ji}$). 

One measure that satisfies these conditions is:

\begin{equation}
	R_{ij}=|ln(p_{ij}) - ln(p_{ji})|
	\label{eq:eq1}
\end{equation}

With,

\begin{equation}
	p_{ij}=\frac{w_{ij}}{w_{i+}}
	\label{eq:eq2}
\end{equation}

Where $w_{ij}$ is the raw weight corresponding to the directed $i \rightarrow j$ arc, and $w_{i+}$ is the strength of the $i^{th}$ vertex as given by \citet{barrat_etal04}:

\begin{equation}
	w_{i+}=\sum_{j \in N(i)}w_{ij}
	\label{eq:eq3}
\end{equation}

Where $N(i)$ is the set of vertices that lie in $i$'s neighborhood (i.e. are connected to $i$ via an outgoing directed arc).  In the case of a social network where the weights are given by the number of communications directed from one actor to another, the strength of each vertex ($w_{.+}$) can be defined as the actors \emph{communicative propensity}.  This is the likelihood that at any given moment a given actor will be active or ``on'', which in our case means being the initiator of a communication event.  We should expect that in human communication networks there should exist substantial heterogeneity across vertices in communicative propensity---with some persons being constantly active, and others communicating more sparingly---which is a phenomenon that is characteristic of other physical and biological systems \citep{barrat_etal04, barthelemy_etal03, serrano_etal09}.

Note that the ``normalized weight'' $p_{ij}$ \citep{serrano_etal09} is the instantaneous probability that if $i$ makes a communication attempt it will be directed towards $j$ (and viceversa for $p_{ji}$); as a probability their sum across $j \in N(i)$ is constrained to be 1.  A substantive interpretation of a reciprocity measure based on the ratio of normalized weights is that a dyad is reciprocal when two persons have the same probability of communicating with one another, and a dyad is non-reciprocal when the probability of one person directing a communication towards another differs substantially from the probability of that person returning that communication (In the following we will just simply call $R_{ij}$ our ``reciprocity'' measure with the caveat that it really stands for the amount of imbalance or non-reciprocity characterizing the dyad).  Factors that affect this probability, such as the number of neighbors connected to each vertex, the relative communicative propensities of each vertex or the dispersion of edge-weights across neighbors for each vertex, should thus be implicated in moving each dyad closer or farther away from the ideal of full reciprocity.  Observe that in the limit if one actor initiates all directed communication attempts while the other actors initiates none, then reciprocity is not defined ($R_{ij}=\infty$), which is consistent with the intuition that there can be no definition of reciprocity when there is no actual \emph{two-way} relationship to speak of.  

\begin{figure*}[t!]
	\centering		
	\includegraphics[width=4in]{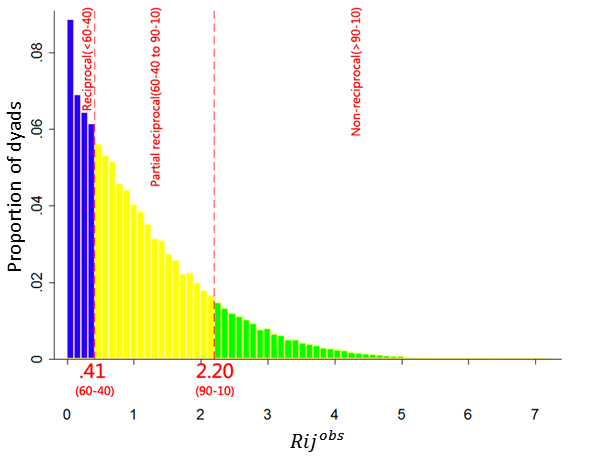}
	\caption{\small Distribution of reciprocity scores across dyads in the cell phone network.  For each edge connecting two vertices, the reciprocity score ($R$) is given by the absolute value of the logged-ratio of the normalized weights ($p_{ij}=w_{ij}/w_{i+}$) corresponding to each directional arc: $R_{ij}=R_{ji}=|ln(p_{ij}/p_{ji)}|$.  $R=0$ indicates full reciprocity.}
	\label{fig:recip-dist}
\end{figure*}

\subsection{Some special cases}
The characterization of reciprocity given above allows us to outline some idealized conditions under which we should expect full reciprocity and under which we should expect systematic deviations from the reciprocity ideal. To build some intuition it helps to rewrite equation~\ref{eq:eq1} as:

\begin{equation}
	R_{ij}=\left| ln \left[ \frac{w_{ij}}{w_{ji}} \frac{w_{j+}}{w_{i+}} \right] \right|
	\label{eq:eq4}
\end{equation}

The first idealized condition that we can consider is an \emph{equidispersion regime} \citep{serrano_etal09, barthelemy_etal05}.  Under this condition, persons distribute their communicative activity equally across partners, with the only constraint being the number of partners ($k_i^{out}$) and their communicative propensity ($w_{i+}$).  It is easy to appreciate that under this regime  the expected directed weights are given by:

\begin{equation}
	\widehat{w}_{ij}=\frac{w_{i+}}{k_i^{out}}
	\label{eq:eq5}
\end{equation}

Substituting~\ref{eq:eq5} into~\ref{eq:eq2} we find that the expected $p_{ij}$ under this regime is simply:

\begin{equation}
	\widehat{p}_{ij}=\frac{1}{k_i^{out}}
	\label{eq:eq6}
\end{equation}

Indicating a strong trade-off between the normalized outflow and the range of contacts for each vertex \citep{aral_alstyne09}. Finally, substituting~\ref{eq:eq6} into~\ref{eq:eq1} shows that in this case the reciprocity equation simplifies to:

\begin{equation}
	\widehat{R}_{ij}=|ln(k_j^{out}) - ln(k_i^{out})|
	\label{eq:eq7}
\end{equation}

Because vertex strength ($w_{.+}$) drops out of the picture under the equidispersion constraint, if persons disperse their calls equally across neighbors, and have the same number of outgoing arcs ($k_i^{out}=k_j^{out}$), then reciprocity is assured. This is the situation depicted in Figure~\ref{fig:config1}.  Thus, when equidispersion holds, deviations from the reciprocity ideal are solely traceable to the magnitude of the degree-differences across the two vertices in a dyad and are \emph{independent of vertex strength differences}.   

A case of non-reciprocity produced by non-assortative mixing by degree is shown in Figure~\ref{fig:config2}.  Here the two vertices match in strength but differ in outdegree.  In this case, even if the two actors were to distribute their communicative activity equally across neighbors they would not be able to reach reciprocity.  The reason for this is that the more sociable green vertex is forced to divide her energy over a larger number of neighbors than the red vertex, reducing the outgoing \emph{probability} of communication in relation to the incoming probability corresponding to her less sociable neighbor \citep{aral_alstyne09}. \emph{This implies that, holding all else equal, degree-assortativity in social networks (the existence of more same-degree dyads than we would expect by chance) should drive the average reciprocity of a random dyad towards the maximum reciprocity point ($R_{ij}=0$)}. Non-assortativity (or negative assortativity) should move dyads towards less reciprocal relations.  

As shown in Figure~\ref{fig:config3}, deviations from the ideal of reciprocity can be produced even when persons share the same number of neighbors and have the same communicative propensities but they do not distribute their communicative activity equally across contacts. In the example shown above, the green vertex follows the equidispersion rule but the red vertex does not. Instead the red vertex concentrates her communicative activity on the green vertex at the expense of her other neighbors.  Setting $w_{i+}=w_{j+}$ in~\ref{eq:eq4}, gives us the expected reciprocity for this case:

\begin{equation}
	R_{ij}=|ln(w_{ij}) - ln(w_{ji})|
	\label{eq:eq8}
\end{equation}

In other words, when vertices have the same strength and have the same number of neighbors, but $R_{ij} \neq 0$, we can be sure that either: (1) at least one of the vertices is investing \emph{more} in that relationship than in his or her other relationships; or (2) at least one of the vertices is investing in that relationship \emph{less} than he or she does in his other relationships. Naturally, both things could be happening at the same time (one partner under-invests while the other one over-invests).

\begin{table*}[t!]
	\centering
	\caption{\small Four variants of weighted reciprocity:  (1) quantity computed in a network with assortative mixing by degree and equal flow dispersion ($\widehat{R}_{ij}^{obs}$); (2) quantity computed in a network with a neutral mixing pattern and equal flow dispersion ($\widehat{R}_{ij}^{rw}$ ; (3) quantity computed in a network with assortative mixing by degree and unequal flow dispersion ($R_{ij}^{obs}$ );(4) ) quantity computed in a network with a neutral mixing pattern and equal flow dispersion ( $R_{ij}^{rw}$ ).}
	\begin{tabular*}{\textwidth}{@{\extracolsep{\fill}} l c c}
		\hline \hline
		&Assortativity&Non-assortativity\\
		\hline
		Equidispersion& $\widehat{R}_{ij}^{obs}$ & $\widehat{R}_{ij}^{rw}$ \\
		Non-equidispersion& $R_{ij}^{obs}$ & $R_{ij}^{rw}$ \\
		\hline
		\label{tab:two-by-two}
	\end{tabular*}
\end{table*}

A fourth case that would produce systematic non-reciprocity according to~\ref{eq:eq4} would be one in which the directed weights for each arc in the dyad match ($w_{ij}=w_{ji}$), but the vertex strength of the partners is different.  In this case, the level of non-reciprocity for that dyad is given by:

\begin{equation}
	R_{ij}=|ln(w_{j+}) - ln(w_{i+})|
	\label{eq:eq9}
\end{equation}

Note that the case of equal weight but non-equal vertex strength is redundant since it is implies that either one partner is under-investing or another partner is over-investing in the relationship; this is therefore another version of the non-equidispersion story shown in~\ref{fig:config3}.  This is intuitive since, as we saw above, when both vertices disperse their communicative activity equally across neighbors $R_{ij}$, is independent of vertex strength differences.  Thus, any dependence of the expected value of $R_{ij}$ on either $w_i+$  or $w_{j+}$ when $w_{ij}=w_{ji}$ can only be produced by deviations from equidispersion.

In a real communication network, we should expect the values of $w_{ij}$ to vary across neighbors for each vertex: equidispersion is an ideal that will usually fail to be met in real social networks \citep{almaas_etal04, barthelemy_etal03}.  Empirical evidence indicates that persons typically divide their neighborhood into core and peripheral members, directing strong (large weight) ties toward core members and keeping only weak (small weight) ties with peripheral members \citep{granovetter73, marsden87}.  Arcs that are considered strong ties in ego's neighborhood should have much larger weights than those that are considered weak ties.  Non-reciprocity results when they are mismatches in the directional tie strength between two vertices: one member of the dyad considers a strong tie what from the point of the view of the other member is a weak tie. \emph{Thus, holding all else equal, deviations from the equidispersion ideal should move the average dyad away from the reciprocity point} ($R_{ij}=0$).  

\begin{figure*}[ht!]
	\centering
	\includegraphics[width=4in]{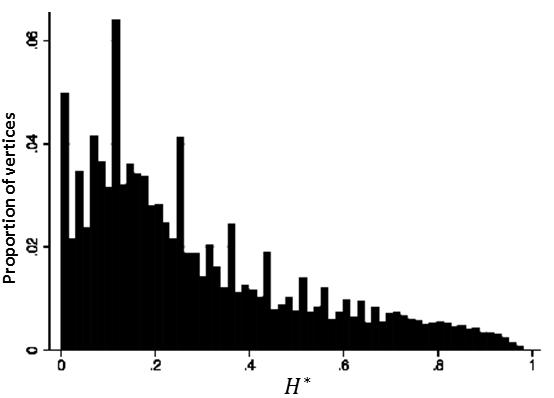}
	\caption{\small Distribution of non-equidispersion scores across vertices in the cell-phone network.  For each vertex, non-equidispersion is given by $H=\sum_j p_{ij}^2$ \citep{serrano_etal09}, which is equivalent to Herfindahl's \citeyearpar{herfindahl50} concentration index.  We normalize this measure to remove any dependence between the expected minimum and vertex degree by   computing $H^*=\frac{H-1/k_i^{out}}{1-1/k_i^{out}}$.  A score of one indicates maximum concentration of communicative activity on one neighbor, a score of zero indicates equidispersion.}
	\label{fig:h-index}
\end{figure*}

\section{The empirical distribution of weighted reciprocity}
\label{sec:empirical}

The data that we will consider in what follows consist of a weighted graph of a human communication network constructed from trace-logs of over 1 billion cellular telephone voice calls made by 8 million subscribers of a single cellular telephone provided in a European country over a two-month period in 2008.  Among these 8 million subscribers there are over 34 million directed arcs, that is instances in which a subscriber made at least one call to the other subscriber.  Of these 34 million arcs, about 16.8 million (49\%) are asymmetric dyads, meaning that the directed arc is not reciprocated.  The remaining 17.2 million symmetric arcs are in 8.6 million mutual dyads consisting of two arcs, indicating that each person in the dyad made at least to the other person during this time period.  The focus below is on these 8.6 million mutual dyads given that reciprocity is only defined for these types of dyads.  We define the weight ($w_{ij}$) of the incoming and outgoing arcs for each vertex as the \emph{number of calls} either received from or made to each neighbor (respectively) during the time period in question. 

Figure ~\ref{fig:recip-dist} depicts the observed distribution of reciprocity computed according to equation~\ref{eq:eq1}.   We divide the observed dyads into three classes: \emph{reciprocal dyads} are those in which the communication probability ratio (taking the largest probability as the numerator) ranges from 1.0 to 1.5 ( (0 to .41 when taking the natural log of the probability ratio).  \emph{Partially reciprocal dyads} are those in which the communication probability ratio is larger than 1.5 but smaller than 9.0  (.41 to 2.20 on the logged scale).  Finally, \emph{non-reciprocal} dyads are those with a probability ratio exceeding 9.0  (2.20 on the logged scale).  We find that a substantial minority (28\%) of dyads belong to the reciprocal class, about 58\% of dyads can be considered partially reciprocal, and a non-trivial minority of dyads (14\%) exhibit extreme non-reciprocity, with one partner being more than nine times more likely to contact the other than being contacted by that partner.  

It is clear that a substantial proportion of dyads in the observed social network  feature relatively large degrees of weighted non-reciprocity. Had we confined ourselves to the purely binary definition of reciprocity as mutuality or symmetry, we would have missed the large levels of communicative imbalance encoded in the directed weights.  This result suggests that there are systematic features of human communicative behavior that drive dyads towards non-reciprocity in spite of often noted psychological preferences and normative expectations for balance in human social relationships \citep{gouldner60, newcomb79, hammer85}.  We investigate some of the topological and structural factors that push social networks either towards and away from reciprocity in what follows.

\subsection{Comparing the observed distribution to alternative regimes}
To what extent are the patterns of reciprocity observed in this social network deviations from what we would expect by chance? To answer this question we compare the observed reciprocity distribution to that obtained from three-alternative regimes, corresponding to three out of the four different configurations in a two-dimensional space defined by the presence or absence of degree assortativity \citep{park_newman03, newman_park03, newman03, newman02, catanzaro_etal04}, versus the presence or absence of a tendency toward equidispersion in the weight distribution of the arcs emanating from each vertex \citep{barthelemy_etal05, barthelemy_etal03, serrano_etal09}.  This is shown in shown in Table~\ref{tab:two-by-two}.  

\begin{figure*}[t!]
	\centering
	\subfloat[\scriptsize Local configuration before rewiring]{
		\includegraphics[width=2.15in]{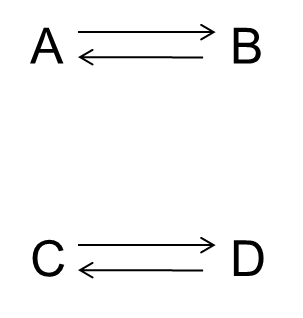}
		\label{fig:before-rewire}
		}
	\subfloat[\scriptsize Local configuration after rewiring]{
		\includegraphics[width=2.15in]{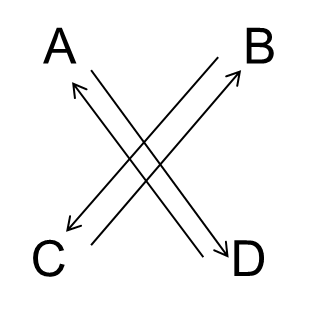}
		\label{fig:after-rewire}
		}
	\label{fig:rewire}
	\caption{Schematic illustration of link-rewiring operation implemented in the Maslov-Sneppen algorithm.}
\end{figure*}

As we have already noted, the observed social network is located in the lower-left corner of the table ($R_{ij}^{obs}$). This is a network displaying positive degree assortativity and a tendency for non-negligible proportion of actors to distribute their communicative activity inequitably across neighbors.  The Pearson correlation coefficient ($r$) between the (excess) degree sequences of each of the two vertices across linked dyads in the observed network is positive:  $r^{obs}_{k_ik_j}=0.33$, which is a value typical for human social networks \citep{newman_park03}.  In addition, as shown in Figure~\ref{fig:h-index}, in the social network that we examine---a substantial number of vertices (among those whose $k_i^{out} \geq 2$) concentrate their communicative outflow on a minority of their contacts,  generating non-equidispersion of weights across the arcs directed at their neighbors.  In the observed communication network, about 10\% of vertices have an non-equidispersion score of  0.66 or above (with 1.0 indicating the extreme case of concentrating all communicative activity on a single partner) and 25\% have a score of .43 or higher.  Only about 25\% of actors come close to the equidispersion ideal ($H^* \leq 0.10$).  The systematic prevalence of substantial inequalities in directed weights across vertices is a property that this social system shares with other physical and biological networked structures \citep[e.g.][]{almaas_etal04, barthelemy_etal03, csermely04, csermely06}.

\subsection{Procedure}

We proceed to generate three alternative comparison networks, all of which preserve the most relevant topological and statistical features of the original network (number of vertices the number of links and the degree-distribution), but which either remove assortativity, impose equidispersion in the distribution of directed weights across neighbors for all vertices, or do both. We remove assortativity in the original network using the Maslov-Sneppen local rewiring algorithm \citep{maslov_sneppen02, maslov_sneppen04} illustrated in Figures~\ref{fig:before-rewire} and \ref{fig:after-rewire}.  It is easy to verify that this procedure preserves the number of edges attached to each vertex, but makes the vertex-to-vertex connections independent of degree. We can verify that the algorithm is successful by computing the degree-correlation after reshuffling.  The resulting network is indeed non-assortative ($r^{rw}_{k_ik_j}=0$) indicative of a ``neutral'' mixing pattern.

Accordingly, the assortative-equidispersed network (upper-left corner) is just like the originally observed network, except that now the number of calls across partners are redistributed and forced to be same ($p_{ij}=1/k^{out}_i$ for all arcs and $R_{ij}=R_{ji}=|ln(k^{out}_j/k^{out}_i)|$ for all dyads); here reciprocity is given by $\widehat{R}_{ij}^{obs})$.  The non-assortative, equidispersed network (upper-right corner of table~\ref{tab:two-by-two}) is just like this last network, except that now the links are reshuffled to remove degree-assortativity according to the procedure described above; here reciprocity is given by $\widehat{R}_{ij}^{rw})$. Finally, the non-assortative non-equidispersed network (lower-right hand corner) is just like this last network, except that the distribution of calls across neighbors matches that of the original data set; here reciprocity is given by $R_{ij}^{rw}$.

Because assortativity and non-equidispersion pull in different directions with respect to reciprocity, we should observe that $\widehat{R}_{ij}^{obs} < R_{ij}^{obs}$ due to the non-equidispersion effect; that is reciprocity in the observed network  (where there is non-equidispersion) is farther away from zero than in a network with similar characteristics but where persons distributed calls equally across partners.   We should also observe that $R_{ij}^{obs} < R_{ij}^{rw}$ due to the assortativity effect; that is reciprocity in the observed network  (which is degree-assortative) is closer to zero than in a network with the same characteristics because, as noted above, reciprocity is more likely among dyads with degree similar nodes.  Finally, due to the non-equidispersion effect, we should expect that $\widehat{R}_{ij}^{rw} < R_{ij}^{rw}$.  That is even in a network without assortativity, one in which persons distribute calls equally across neighbors should have reciprocity values closer to zero than one where this condition does not obtain. If these three inequalities hold, then we should find the following partial ordering of expected (average) non-reciprocity across the four networks:

\begin{equation}
	\widehat{R}_{ij}^{obs} < min\left(\widehat{R}_{ij}^{rw}, R_{ij}^{obs} \right) \leq  max\left( \widehat{R}_{ij}^{rw}, R_{ij}^{obs} \right)  < R_{ij}^{rw}
\end{equation}

The most reciprocal network should be the one which has both assortativity and equidispersion, and the least reciprocal network should be one without assortativity and without equidispersion.  Note that the ordering of the expected values of $\widehat{R}_{ij}^{rw}$ and $R_{ij}^{obs}$ cannot be predicted a priori, since the question of which force is greater, (1) the ability of assortativity to drive reciprocity towards zero or (2) the ability of non-equidispersion to move the same quantity away from zero, is an empirical issue. We can however expect that reciprocity in these two networks should fall in between the two extremes described above, since they are positive in a factor that lowers reciprocity and negative on a factor that increases it. If assortativity is a stronger factor in driving non-reciprocity towards zero than non-equidispersion is in driving it away from zero, then we should find that $\widehat{R}_{ij}^{rw} > R_{ij}^{obs}$.  If the opposite is the case, then we should find that $\widehat{R}_{ij}^{rw} < R_{ij}^{obs}$.

\begin{figure*}[ht!]
	\centering
	\includegraphics[width=4.5in]{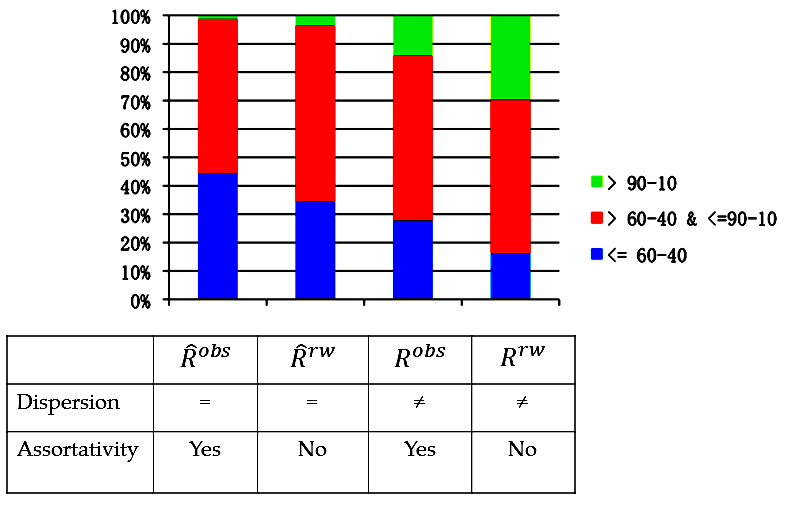}
	\caption{Distribution of dyadic reciprocity in the observed cell phone network and three artificial variations.  Blue bars are \emph{reciprocal dyads}, red bars are \emph{partially reciprocal dyads} and green bars are  \emph{non-reciprocal dyads} as defined in section~\ref{sec:empirical} above. }
	\label{fig:money-shot}
\end{figure*}

\subsection{Results}
\label{subsec:results}

Figure~\ref{fig:money-shot} summarizes the differences in the relative distribution of reciprocity across all four networks.  The results largely agree with expectations regarding the two regimes that should fall at the lower and higher extremes of weighted reciprocity.  Thus, the network without assortativity and without equidispersion  ($R^{rw}$) displays proportionally more dyads with  extreme levels of non-reciprocity. While only about 14\% of dyads in the observed network ($R^{obs}$) exhibit extreme non-reciprocity (e.g. one partner being nine times more likely to initiate a communication attempt than than the other), this proportion more than doubles once we remove the assortativity bias but keep everything the same (30\%). Meanwhile while about 28\% of dyads enjoy some level of reciprocity in the observed network, this number drops to 16\% in the non-assortative version of the same network.  Also as expected, the network displaying reciprocity values closest to the zero (full reciprocity) level is the one that has both assortativity and equidispersion. Here the proportion of reciprocal dyads is 44\% (in comparison to 28\% in the original data), and the proportion of extremely non-reciprocal dyads is only 1\%. 

The results shown in figure~\ref{fig:money-shot} provide an answer to the question of which of the two tendencies observed in human communication networks---assortativity or non-equidispersion---contributes more to system level reciprocity.  The answer is clear: adding equidispersion to the least reciprocal network results in a much more dramatic move towards reciprocity than does adding assortativity to the same network  (compare the difference between $R^{rw}$ and $\widehat{R}^{obs}$ to the difference between  $R^{rw}$ and $R^{obs}$). In this respect, while assortativity keeps human communication networks from resembling the least reciprocal of our baseline networks, the tendency to disperse communication activity inequitably across contacts is responsible for the bulk of the observed non-reciprocity.  

Accordingly, the final ordering of expected reciprocity (with smaller values indicating more reciprocity) for all the four networks is as follows:

\begin{quotation}
	$\widehat{R}_{ij}^{obs} < \widehat{R}_{ij}^{rw} < R_{ij}^{obs} < R_{ij}^{rw}$
\end{quotation}

\section{Discussion}
\label{sec:discussion}

In this paper we have defined a metric for reciprocity applicable to weighted networks.  Under this conceptualization, reciprocity is defined as balance in the number of communications flowing from one partner to another, normalized by the communicative activity of each person. This yields a notion of reciprocity that is interpretable as a \emph{matching} of the \emph{probabilities} that the two vertices in a dyad will initiate directed contact attempts towards each other.   When persons match in overall communicative propensity ($w_{.+}$), reciprocity reduces to the (absolute value of the logged) ratio of the weights of the incoming and outgoing arcs. When the weights ($w_{ij}$, $w_{ji}$) of the arcs are the same, reciprocity simplifies to the (absolute value of the logged) ratio of the strength of the vertices.  The most revealing special case results when vertices disperse their communication attempts equally across neighbors.  In this case reciprocity simplifies to the (absolute value of the logged) ratio of the \emph{number of neighbors} (outdegree) of each vertex.  

We examined the distribution of reciprocity as defined here in a social network built from trace logs of cell-phone communications between individuals during a two month period.  We found that these relationships exhibit varying levels of balance, with the majority of relationships exhibiting moderate to extreme large imbalances.  In this respect, while reciprocity might certainly be a communicative preference across persons, there are systematic features of human communication behavior and network topology that prevent it from becoming a statistical ``norm'' as would be predicted by cognitive balance and normative theories of reciprocity \citep{heider58, gouldner60, newcomb79, hallinan78}.  One such feature consists precisely of the propensity  to divide contacts into strong and weak ties, thus concentrating communicative activity on a few partners at the expense of others.  As we have shown, eliminating this tendency---by imposing equidispersion of weights on the observed network---moves it closer to the ideal of full-reciprocity.  In addition, we demonstrate that one systematic feature that differentiates social networks from other networks---namely, the tendency of like to associate with like as manifested in the degree-assortativity property \citep{park_newman03}---makes the observed communication network more reciprocal.  Assortative mixing creates reciprocity by facilitating the matching of probabilities across incoming and outgoing arcs (see Figure~\ref{fig:net-configs}).  When we remove assortativity from the observed network, we observe that non-reciprocity increases dramatically.  It is worth noting, however, that the effect of non-equidispersion in moving reciprocity away from the ideal of zero is stronger than the effect of assortativity in moving this quantity closer to zero.

These results have important implications for how we think about the phenomenon of dyadic reprocicity in social and other networked systems.  For instance, smaller groups or dense communities that impose homogeneity in most topological characteristics (including the number of neighbors as in fully-connected cliques) should exhibit more weighted reciprocity than social systems that induce large inequalities in connectivity across partners (e.g., social systems characterized by ``popularity tournament'' dynamics) \citep{martin09, waller37, gould02, barabasi_albert99}.  Networked systems that induce anti-correlation in the number of neighbors of each vertex in a dyad should---all else being equal---be characterized by high-levels of non-reciprocity.  In the same way, positive correlations across vertices on other relevant characteristics (e.g. average outgoing arc weight or vertex strength) should move social relationships towards the reciprocity ideal, while mismatches in these vertex-level traits should increase non-reciprocity. In this respect, observed tendencies for persons to match in these traits may be the indirect result of an underlying tendency to preserve more reciprocal relationships available in the network and terminate the least reciprocal---biased selection into reciprocal relationships---than a direct preference to be concordant on these surface features.

\end{document}